\documentstyle[prc,aps,epsfig]{revtex}
\newcommand{\be}{\begin{eqnarray}}
\newcommand{\ee}{\end{eqnarray}}

\newcommand{\bi}{\begin{itemize}}
\newcommand{\ei}{\end{itemize}}
\def\lsim{\mathrel{\rlap{
\lower4pt\hbox{\hskip-3pt$\sim$}}
    \raise1pt\hbox{$<$}}}     
\def\gsim{\mathrel{\rlap{
\lower4pt\hbox{\hskip-3pt$\sim$}}
    \raise1pt\hbox{$>$}}}     

\begin{document}
\twocolumn[\hsize\textwidth\columnwidth\hsize
           \csname @twocolumnfalse\endcsname

 \title{Anomalous radial expansion in central heavy-ion 
reactions}
\author{K. Morawetz$^{1,2}$, M. P{\l o}szajczak$^2$ and
V.D. Toneev$^{2,3}$}
\address{$^1$ LPC-ISMRA, Bld Marechal Juin, 14050 Caen, 
France\\
$^2$ Grand Acc\'{e}l\'{e}rateur National d'Ions Lourds 
(GANIL),\\
CEA/DSM -- CNRS/IN2P3, BP 5027, F-14076 Caen Cedex 05, 
France \\
$^3$ Bogoliubov Laboratory of Theoretical Physics,  
Joint Institute for Nuclear Research, \\ 141980 Dubna, 
Russia }
\maketitle

\begin{abstract}
The expansion velocity profile in central heavy-ion 
reactions
in the Fermi energy domain is examined. The radial 
expansion is non-Hubblean and in the surface region it 
scales proportional 
to a higher exponent ($\alpha > 1$) of the radius. 
The anomalous expansion velocity profile is accompanied 
by a power law 
nucleon density profile in the surface region. Both 
these features of central 
heavy-ion reactions disappear at higher energies, and 
the system 
follows a uniform Hubble expansion ($\alpha \simeq 1$). 
\end{abstract}
\vskip2pc]

\section{Introduction}
In heavy-ion collisions, the stage of 
compression and heating is followed by the expansion 
of nuclear matter. Expansion dynamics as a collective 
motion of excited matter is characterized by certain 
space-momentum correlations and has
been well  ascertained by experimental data such as the 
collective flow. 
In particular, in central heavy-ion collisions
with the beam energy ranging from  the Fermi energy to 
almost 200 GeV/A, the radial flow  is clearly manifested 
through the 
flattening of the transverse spectra with the 
particle mass and this effect, as expected, is 
stronger for heavier systems \cite{Dan99}. 

The collective expansion scenario is important also in
other issues. In studying the quantum statistical 
correlations
which describe the space-time characteristics of 
expanding systems, one can infer the information about the 
freeze-out configuration. 
The recent finding here is that
the size parameters of an effective source are 
determined not only by the
geometrical length scale which measures the region of 
homogeneity \cite{MS88} but also by the thermal length 
scale which is related 
to the region in the coordinate space from which 
identical particles with 
similar momenta may emerge \cite{CLZ94}. Alongside with 
the temperature and the freeze-out
time, the thermal size of a source is determined by 
the velocity gradients in this source. The thermal 
length  
 dominates the correlation function if the geometrical 
length scale is 
sufficiently large \cite{CLZ94}. Similarly, the question 
about 
the particle velocity profile at the freeze-out 
configuration 
arises when one considers the formation of light
nuclei in the framework of a coalescence model 
\cite{PMB98}.  

In almost all papers devoted to this subject, the 
velocity profile 
 is assumed to be linear :
\be \label{eq1}
\vec v = \frac{\dot R(t)}{R(t)} \ \vec r 
\ee
where $\vec v$ and $\vec r$ are three- or 
two-dimensional  vectors for the
spherically or cylindrically symmetric 
expansion, respectively.  The relation (\ref{eq1})  is
prompted by an analytical scaling solution of the 
equations of non-relativistic 
hydrodynamics with the ideal gas equation of state for 
a slowly expanding fireball \cite{BGZ78,CCL98}. In fact, 
Eq. (\ref{eq1}) is a consequence of a
regular motion governed by  the continuity equation when 
a fluid does not
influence the expansion rate. This relation  
is well known in cosmology where the Hubble constant 
\cite{H73,MTW72} :
$\dot R(t)/R(t) \approx  65$ km/s/Mpc, 
at the right-hand side of Eq. (\ref{eq1}), characterizes 
the expansion of homogeneous and isotropic galaxies 
\cite{MTW72}. 

It is important to note that  Eq. (\ref{eq1}) need not 
be 
valid in general. As was shown by Dumitru \cite{D99}, 
while
the longitudinal expansion of the fireball three-volume 
is 
independent of the energy density of
the fluid, the transverse collective motion in the case 
of 
(3+1)-dimensional expansion  may couple the expansion 
rate to the properties of the fluid, {\it i.e.}, to the 
equation of state.  In 
particular, the hydrodynamic
solution for a fireball expanding in the longitudinal 
and transverse direction
with a possible first order hadronization phase 
transition affects the three-volume expansion rate on 
the hadronization 
hypersurface \cite{D99}.  

Analyzing the experimental data on nuclear 
multifragmentation
within an extended statistical microcanonical model 
which takes into account an interplay of  the radial 
expansion with a 
non-spherical shape of the fragmenting nucleus 
\cite{LPT99}, 
it turned out to be necessary to postulate a 
non-Hubblean
velocity profile $\vec v(\vec r)$  for fragments in the 
freeze-out volume :
\be \label{eq2}
\vec v(\vec r) = v_0 \left( \frac{r}{R_0} 
\right)^{\alpha} \ \frac{\vec r}{r}
\ee
with $\alpha$ in between 1.5 and 2 \cite{LPT99}, to 
explain 
the experimental charge-number dependence of 
the mean kinetic energy of
fragments in central $Xe+Sn$ collisions at 50 MeV/A.
The answer to the question : why the exponent  $\alpha$ 
of the radial expansion
differs from 1 ($\alpha=1$ yields the Hubble expansion), 
cannot be given within the statistical 
multifragmentation model.  
An application 
of the hydrodynamics for this kind of problem is also 
questionable 
because the nucleon density at the freeze-out 
configuration 
is low and, moreover, the dynamical
processes at short time scales are not correctly 
described. 
For that reason, in this work we
study both the particle velocity profile and the 
particle density profile 
in central HI collisions using the framework of a 
nonlocal quantum 
kinetic theory \cite{SLM96}.
To gain an insight into the dynamics of collective 
expansion of small fermionic
systems such as the atomic nuclei, the kinetic approach
which uses the quasiparticle interaction as input and
takes into account consistently the two-particle 
correlations \cite{SLM96} is
probably more reliable, even though the
dynamical formation of clusters is absent in this 
approach. 

The paper is organized as follows. In Sect. II.A, the 
main ingredients of the
nonlocal quantum kinetic approach are presented. The 
time evolution of central
$Ta+Au$ collisions at 33 MeV/A and 60 MeV/A is studied 
in Sect. II.B 
by looking at the transversal and longitudinal profiles 
of 
the nucleon velocity, the nucleon density and the proton 
to neutron ratio. The
expansion velocity profile is discussed in more details 
in Sect. II.C
, separately for bulk and surface particles. The 
qualitative evolution of the
radial expansion profile with the collision energy is 
compared in Sect. II.D 
with the dynamical trajectories of excited system in the 
temperature - particle density plane. The possible 
consequences of the
long-range tail in the particle density on the small 
momenta behavior of the
Bose-Einstein correlations in discussed in Sect. II.E. 
Finally, Sect. III 
summarizes main results of the paper.

\section{The kinetic approach}
\subsection{The nonlocal quantum kinetic equation}

 The observables of interest are : the particle density 
$n(r,t)$, the current 
density $J(r,t)$ and the kinetic energy density 
$E(r,t)$, which can
be expressed by the one - particle phase - space 
distribution 
function $f(p,r,t)$ as follows~:
\be
n(r,t)&=&\int {d p\over (2\pi)^3} \ f(p,r,t)\nonumber\\
J(r,t)&=&\int {d p\over (2\pi)^3} \ p \ f(p,r,t) \\
E(r,t)&=&\int {d p\over (2\pi)^3} \ {p^2 \over 2 m} \ 
f(p,r,t) \nonumber ~ \ .
\label{j}
\ee
The one - particle distribution function obeys a 
nonlocal Boltzmann - Uehling - Uhlenbeck (BUU) kinetic 
equation \cite{SLM96} :
\begin{eqnarray}
\!\!\!&&{\partial f_1\over\partial 
t}+{\partial\varepsilon_1\over\partial k}
{\partial f_1\over\partial 
r}-{\partial\varepsilon_1\over\partial r}
{\partial f_1\over\partial k} = 
\sum_b\int{dpdq\over(2\pi)^6} \  {\cal P}
\nonumber\\
&\times&\Bigl[f_3f_4\bigl(1-f_1\bigr)\bigl(1-f_2\bigr)-
\bigl(1-f_3\bigr)\bigl(1-f_4\bigr)f_1f_2\Bigr] ~ \ ,
\label{9}
\end{eqnarray}
with the Enskog-type shifts of the arguments 
\cite{SLM96} :
\begin{eqnarray}
\label{deltas}
f_1 & \equiv & f(k,r,t)  \nonumber \\
f_2 & \equiv & f(p,r\!-\!\Delta_2,t)  \nonumber \\
f_3 & \equiv & 
f(k\!-\!q\!-\!\Delta_K,r\!-\!\Delta_3,t\!-\!\Delta_t) 
 \\
f_4 & \equiv & 
f(p\!+\!q\!-\!\Delta_K,r\!-\!\Delta_4,t\!-\!\Delta_t) ~ 
\ .
\nonumber 
\end{eqnarray}
The arguments of the
effective scattering measure ${\cal P}$ are centered in 
all
$\Delta$ - shifts. The quasiparticle energy 
$\varepsilon$ contains 
the mean field as well as the correlated self energy.
The shifts or displacements are a compact form of 
gradient corrections 
and ensure that the conservation laws contain both the 
mean-field  
and the two-particle correlations. In particular, the 
momentum and the energy 
gain arises from the finite duration of collisions 
\cite{LSM99}. 
All shifts in (\ref{deltas}) are proportional to 
derivatives of 
the scattering phase shift \cite{SLM96,NTL91,H90} 
and have been calculated for realistic nuclear 
potentials \cite{MLSK98}.
When neglecting these shifts, one recovers the usual BUU 
scenario.

It should be noted that using the nonlocal 
BUU kinetic equations \cite{SLM96} 
for the description of the proton spectra in central 
$Xe + Sn$ collisions at 50 MeV/A
leads to a significant enhancement of the 
high energy tail \cite{MLSCN98} and a better agreement
with the experimental data than obtained using 
the standard BUU equations.

\subsection{The evolution plots}
\begin{figure} 
\psfig{file=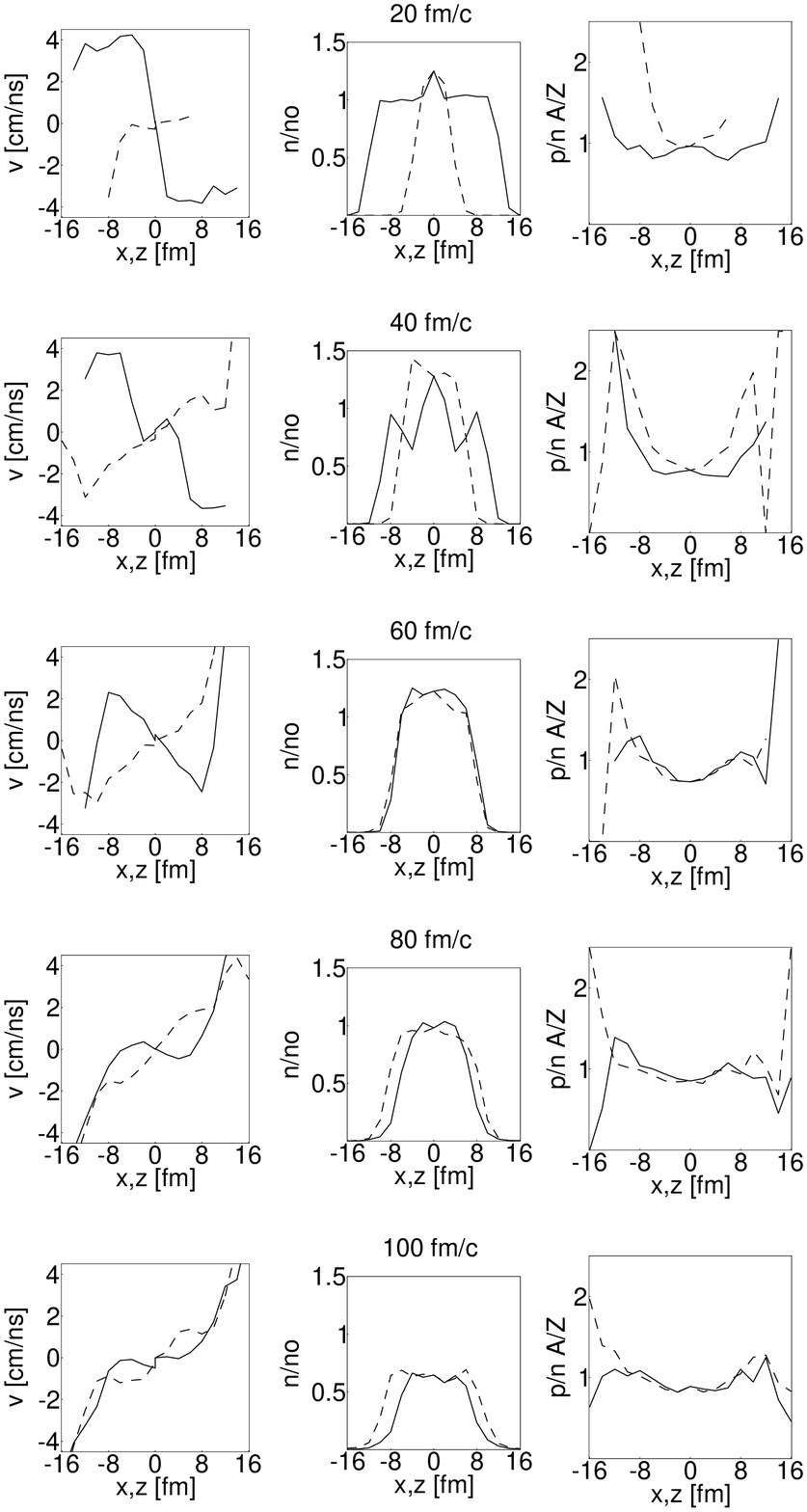,height=17cm,angle=0} 
\vspace{0.5cm} 

\psfig{file=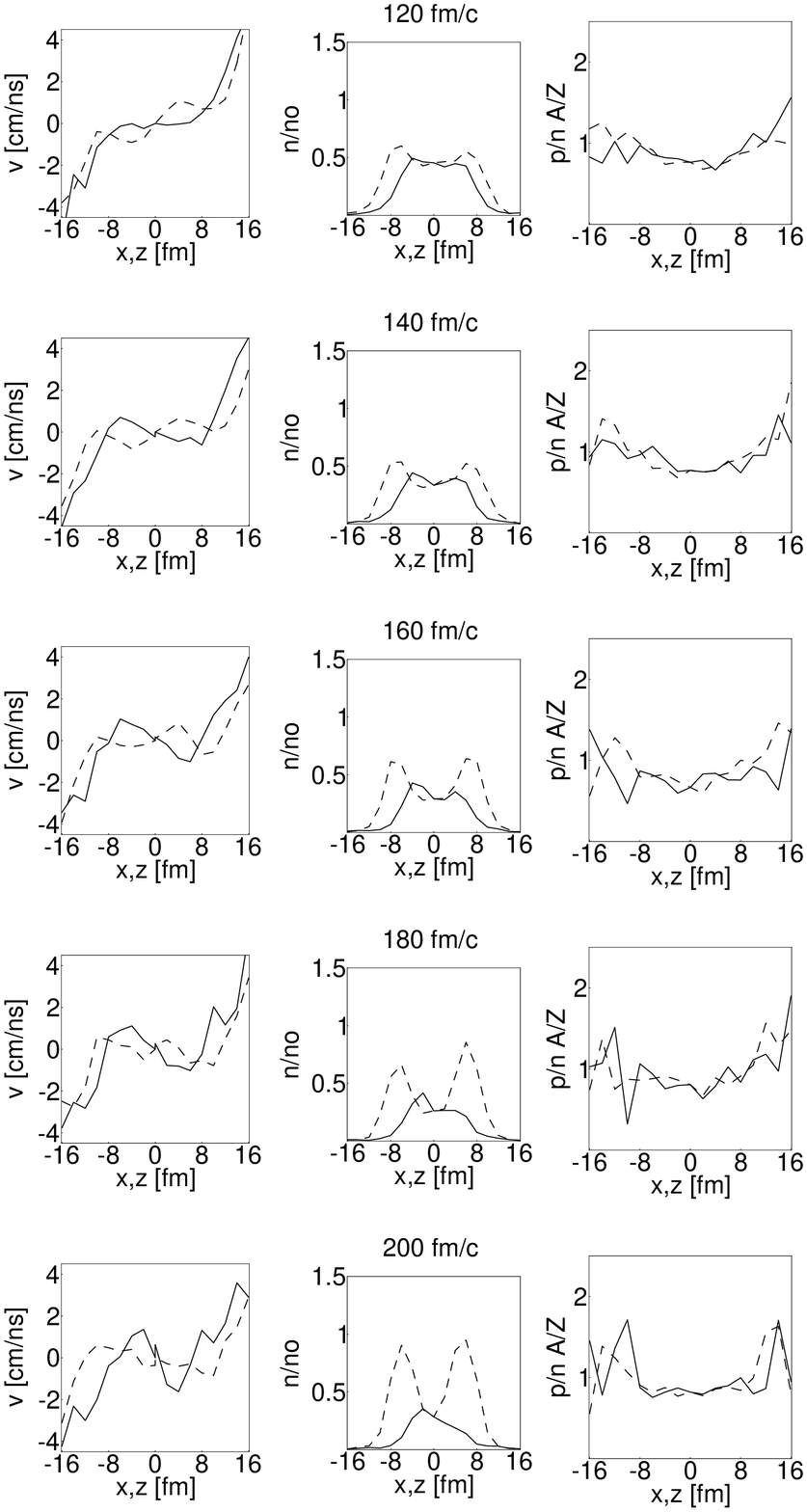,height=17cm,angle=0} 
\caption{The time evolution of central $Ta + Au$ 
collisions at 
$E_{lab}/A = 33$ MeV in  the nonlocal kinetic model 
\protect\cite{LSM99}.   
The first column represents the velocity profile. 
Both transversal (the $x$-direction) (the dashed line) 
and the 
  longitudinal (the $z$-direction) (the solid line) 
profiles are shown.  
The second column shows the longitudinal (the solid 
line) and the
  transversal (the dashed line) density profiles in 
units of $fm^{-3}$. 
The third column presents the ratio of proton and 
neutron densities 
normalized to the initial $Z/A$ - ratio for both
the longitudinal (the solid line) and the transversal 
(the dashed line)
projections.}
\label{fig1} 
\end{figure}

\begin{figure} 
\psfig{file=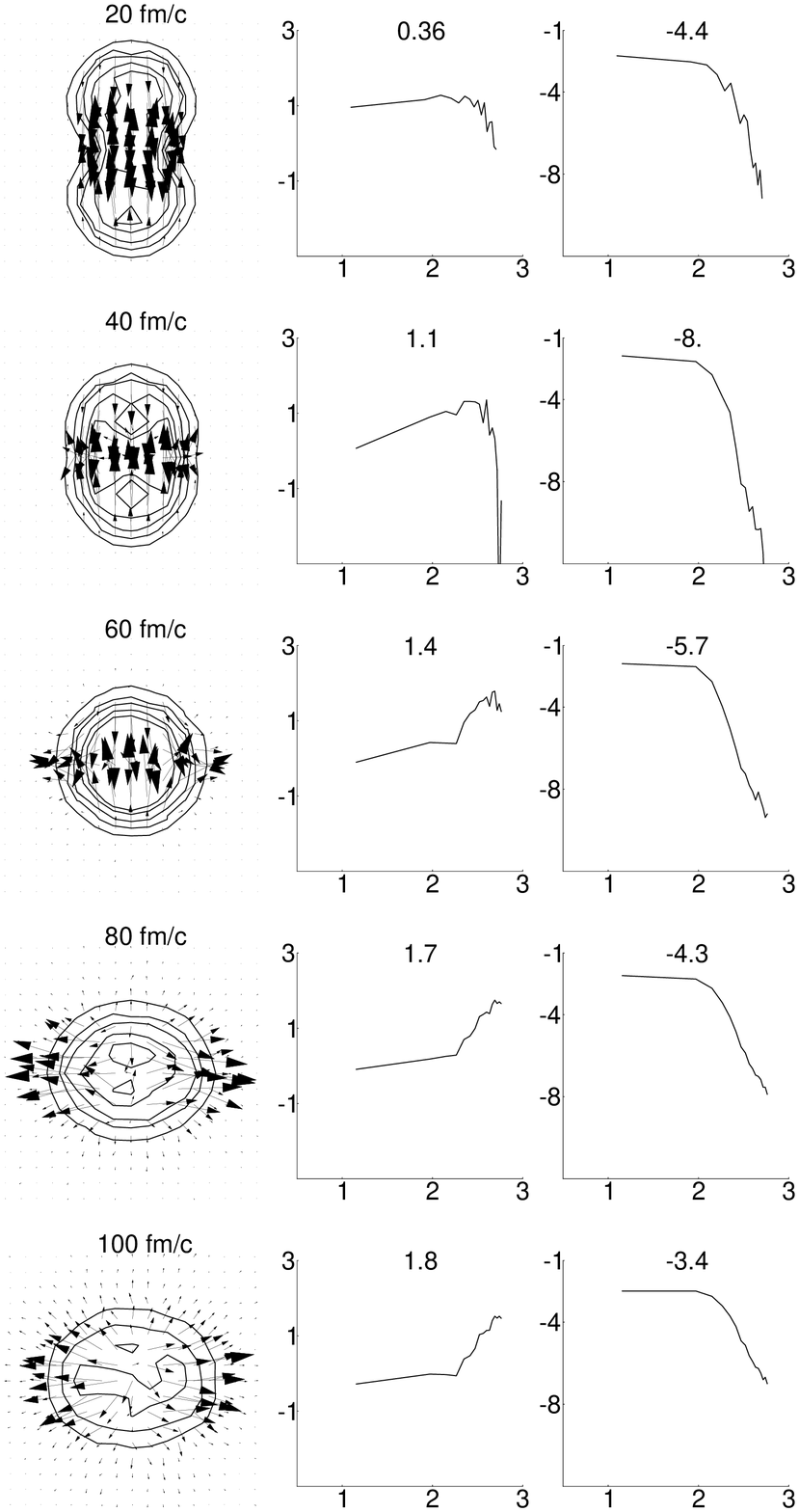,height=17cm,angle=0} 
\vspace{0.5cm} 

\psfig{file=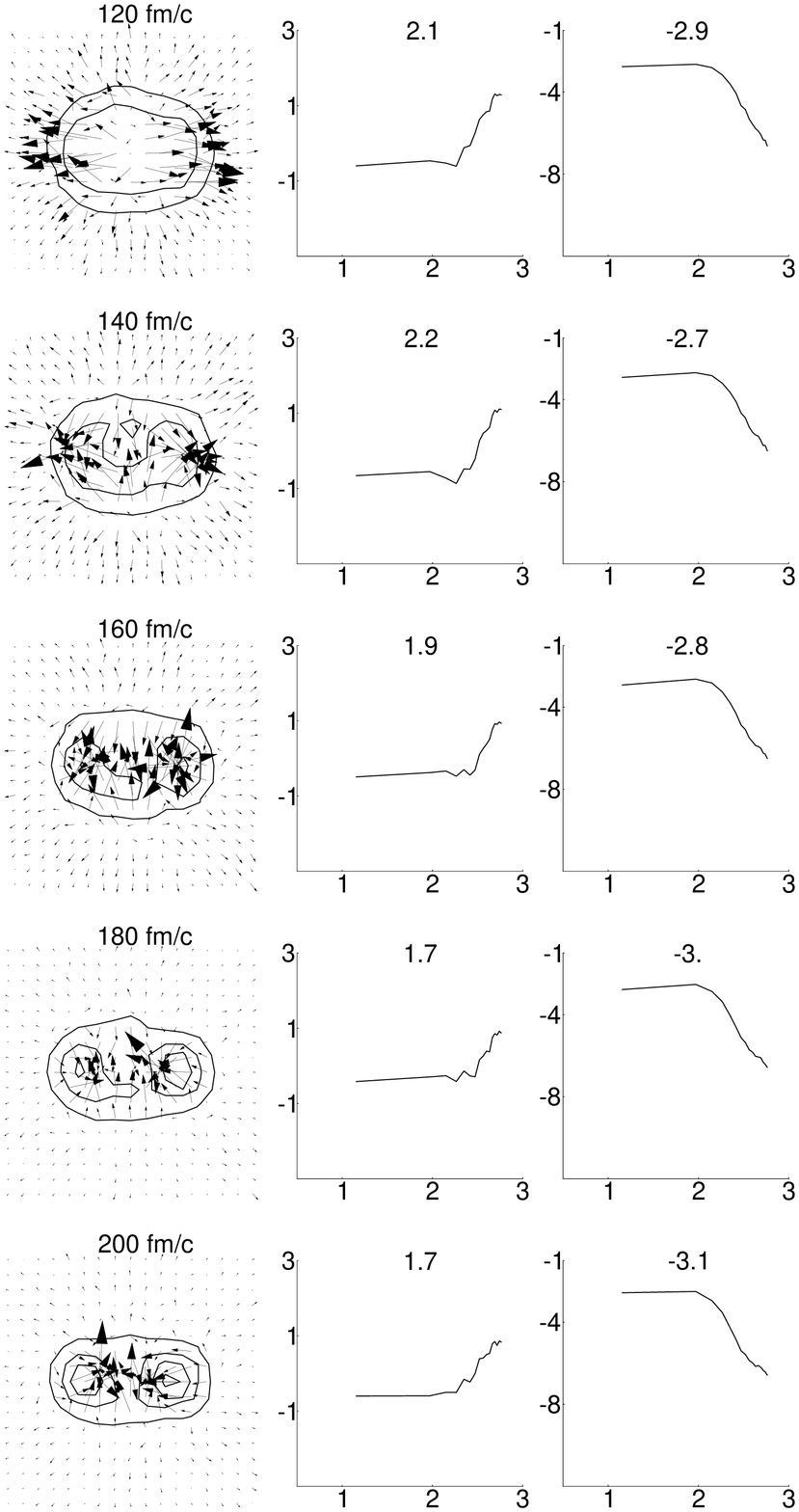,height=17cm,angle=0} 
\caption{The time evolution of central $Ta+Au$ 
collisions at 
$E_{lab}/A = 33$ MeV in  the 
nonlocal kinetic model \protect\cite{LSM99}. Plots 
in the first column show the $(x-z)$ - density cut. 
The mass momenta are shown by arrows. The second column 
shows the log - log  plot of the 
angular averaged modulus of the expansion velocity 
versus the radius. The slope of the straight line fit of 
the 
surface matter expansion profile for  $R > 10$ fm is
indicated at each plot (see also Fig. 
\protect\ref{fig3}). 
The third column shows the log - log plot of 
the angular averaged nucleon density versus the radius. 
The slope of the straight line fit
of the surface profile is indicated at each plot as 
well.
\label{fig1a} }
\end{figure}

The result of the nonlocal BUU scenario for the reaction 
$Ta+Au$ at 33 MeV/A
can be seen in Figs. \ref{fig1} and \ref{fig1a}. 
Let us concentrate first on the corresponding velocity 
and density profiles
(the first and the second column in Fig. \ref{fig1})
and the arrows characterizing the mass momentum (the 
first column in Fig.
\ref{fig1a}). One 
sees that at around 40 fm/c the nuclei start to squeeze 
out the matter
side-wards (the first column in Fig. 2), what is 
characterized by the momentum 
focusing at both sides perpendicular to the beam 
direction 
, predominantly in peripheral regions. The surface 
matter is stopped
and bounced back in longitudinal direction during the 
times
$20 - 80$ fm/c (see the first column in Fig. 
\ref{fig1}). The inner (bulk) 
matter exhibiting a quite clear
spatial boundary (see the second column in Fig. 
\ref{fig1a}) is still moving
inwards. This leads at short time scales ($\sim 40$ 
fm/c) to 
an enhancement of matter density. The strong velocity 
gradient, which is
seen at $60$ fm/c, disappears
at about $100$ fm/c and the inner matter comes to a 
rest. 
The recoil of the splashing matter at
the surface and the attractive mean field force start 
effectively to
reaccelerate an inner matter towards the
center of mass. Since the surface particles are still 
accelerating in the
outward direction, therefore
there is a zone of matter in between which comes to a 
rest. This evolution
leads to the dumb-bell structure  in the transversal 
density profile 
at $120 - 240$ fm/c. 

The different behavior of the surface matter and the 
bulk
matter leads to the development of a nonlinear velocity 
profile in the surface
region, which can be seen in the log - log 
representation of the 
angular averaged velocity in the second column of Fig. 
\ref{fig1a}. 
For $t > 80$ fm/c, the velocity - radius  scaling  with 
an asymptotically 
stable coefficient $\alpha_{surf} \simeq 1.75 \pm 0.05$ 
(see the second column in Fig. \ref{fig1a})
appears definitely in the surface region. In this 
region, 
the particle density drops nearly as a power law : 
$n(r) \sim r^{-\beta}$, with
the asymptotically  stable  (for $t > 120$ fm/c) 
coefficient :
$\beta \simeq 3 \pm 0.2$. As can be seen from the $N/Z$ 
- ratios  
in the third column in Fig. \ref{fig1},
the Coulomb interaction expels protons from the surface
in the early stage of the evolution. In particular, the proton rms
radius is larger than the neutron one. This effect 
becomes weaker at later times ($t \simeq 200$ fm/c) but, nevertheless, 
it survives indicating that proton and neutron
distributions are different in the surface region. This may lead to different
source temperatures and temperature gradients for protons and neutrons and,
hence, to different quantum statistical corrections  for protons and neutrons
in the interferometry experiments. This possibility has been suggested in the
phenomenological analysis of the asymmetric reaction $Ar + Au$ at 30 MeV/A
\cite{HCA97}.
   
These two unusual effects~: the nonlinear velocity 
profile with 
$\alpha_{surf} \in [1.5,2.0]$, 
and an approximately power law fall-off of the particle 
density with
$\beta \simeq 3$, characterize the transitional region 
in the central collisions of symmetric HI collisions in 
the Fermi energy domain.

For later times ($t > 200$ fm/c), one sees the formation 
of an oblate 
configuration which is
connected to the inversion of the velocity of the inner 
matter and to the
accumulation of the density. In agreement with earlier
observations~\cite{BRRS93,Dan95},  we see that 
 the formed hot and nearly fused matter is not 
spherically
symmetric. We would like to remark that this deformation 
is specific
for energies around the Fermi energy and, moreover, is 
impact-parameter dependent.
At  nonzero impact parameters, the shape of the matter 
distribution becomes prolate due to the spectator matter 
keeping its 
initial direction of motion and  also due to the angular 
momentum effects. 

\begin{figure} 
\psfig{file=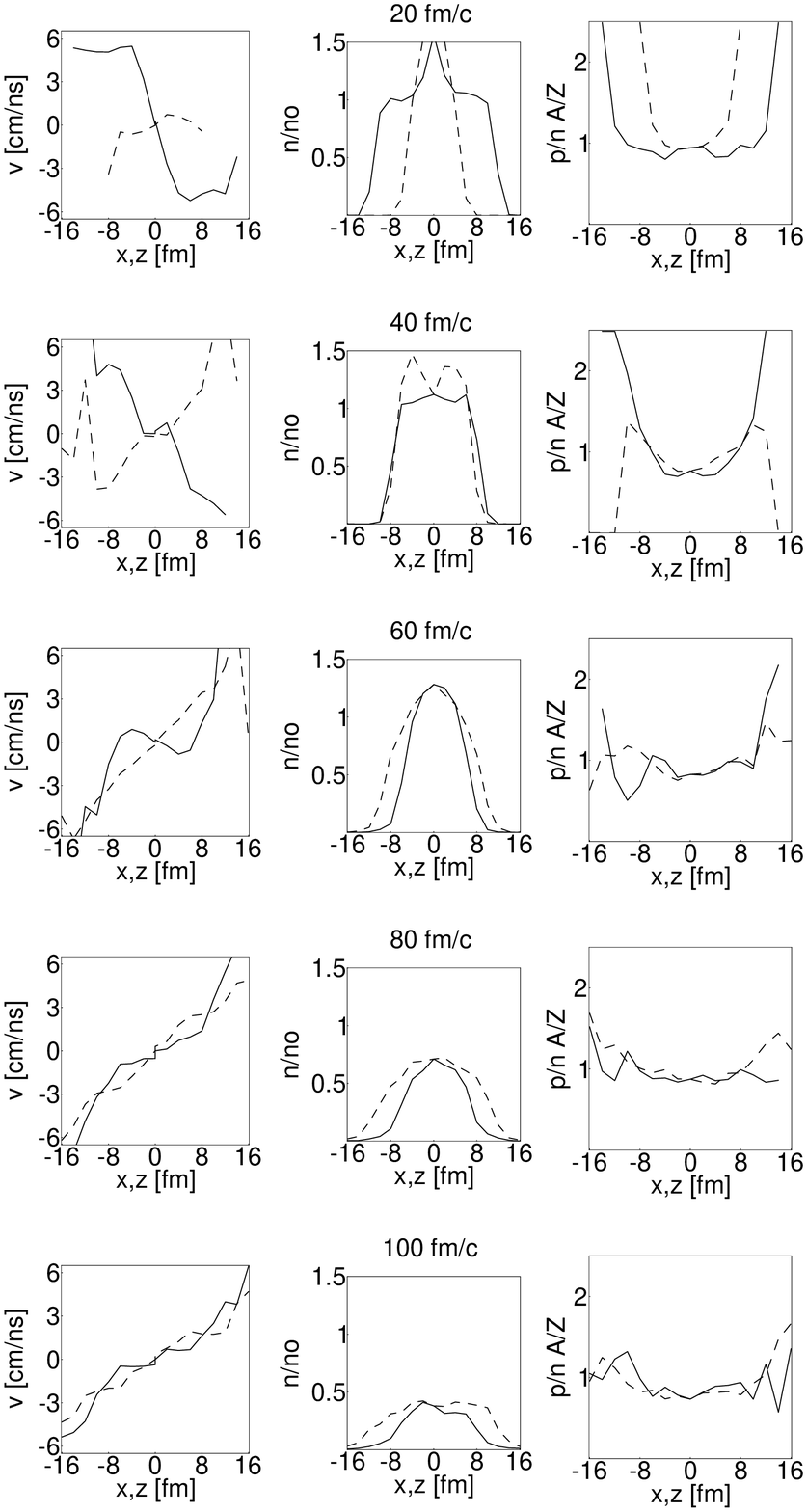,height=17cm,angle=0} 
\vspace{0.5cm} 

\psfig{file=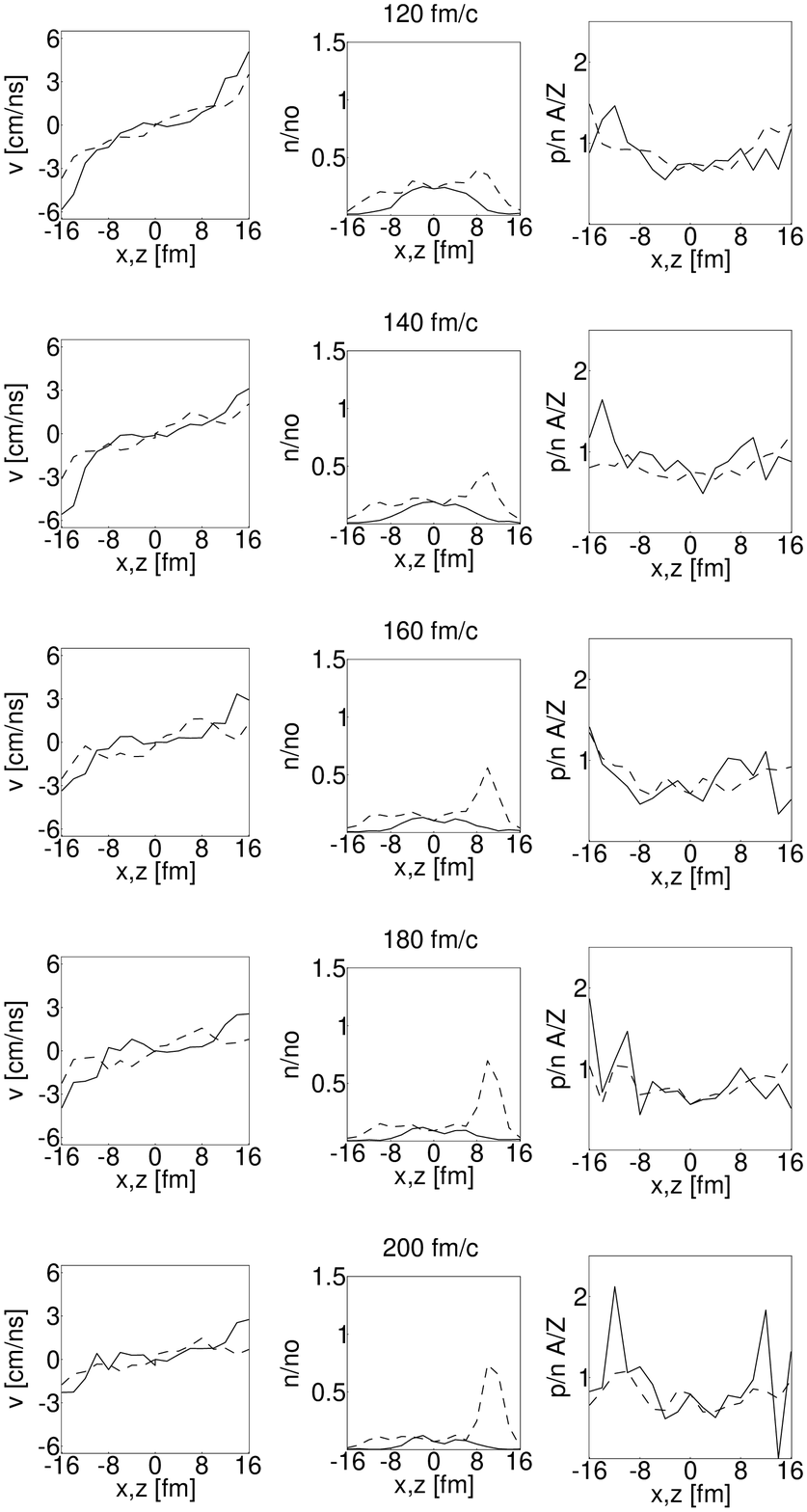,height=17cm,angle=0} 
\caption{The same as in Fig. 1 but for $E_{lab}/A = 60$ 
MeV.}
\label{fig2} 
\end{figure}

\begin{figure} 
\psfig{file=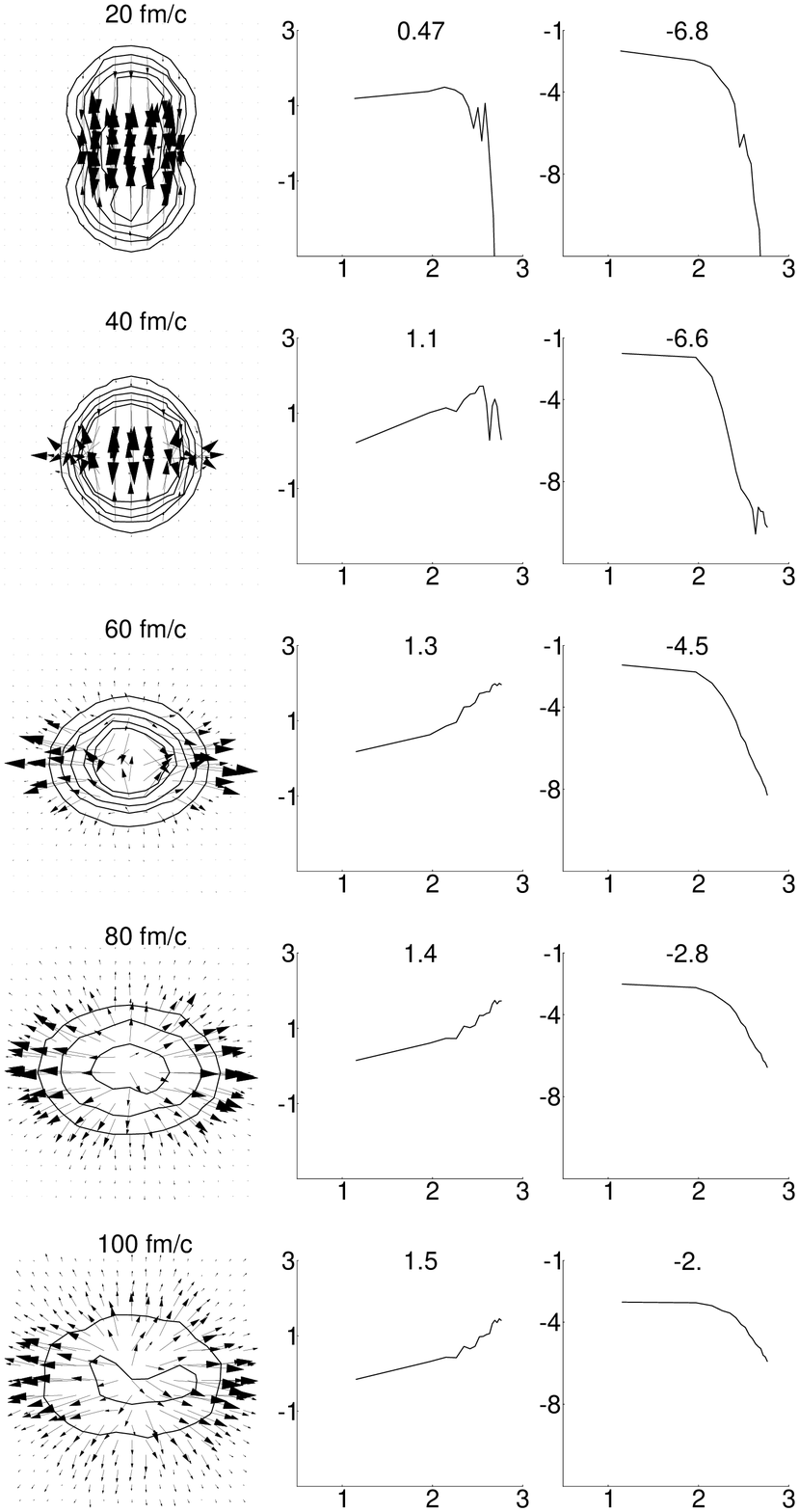,height=17cm,angle=0} 
\vspace{0.5cm} 

\psfig{file=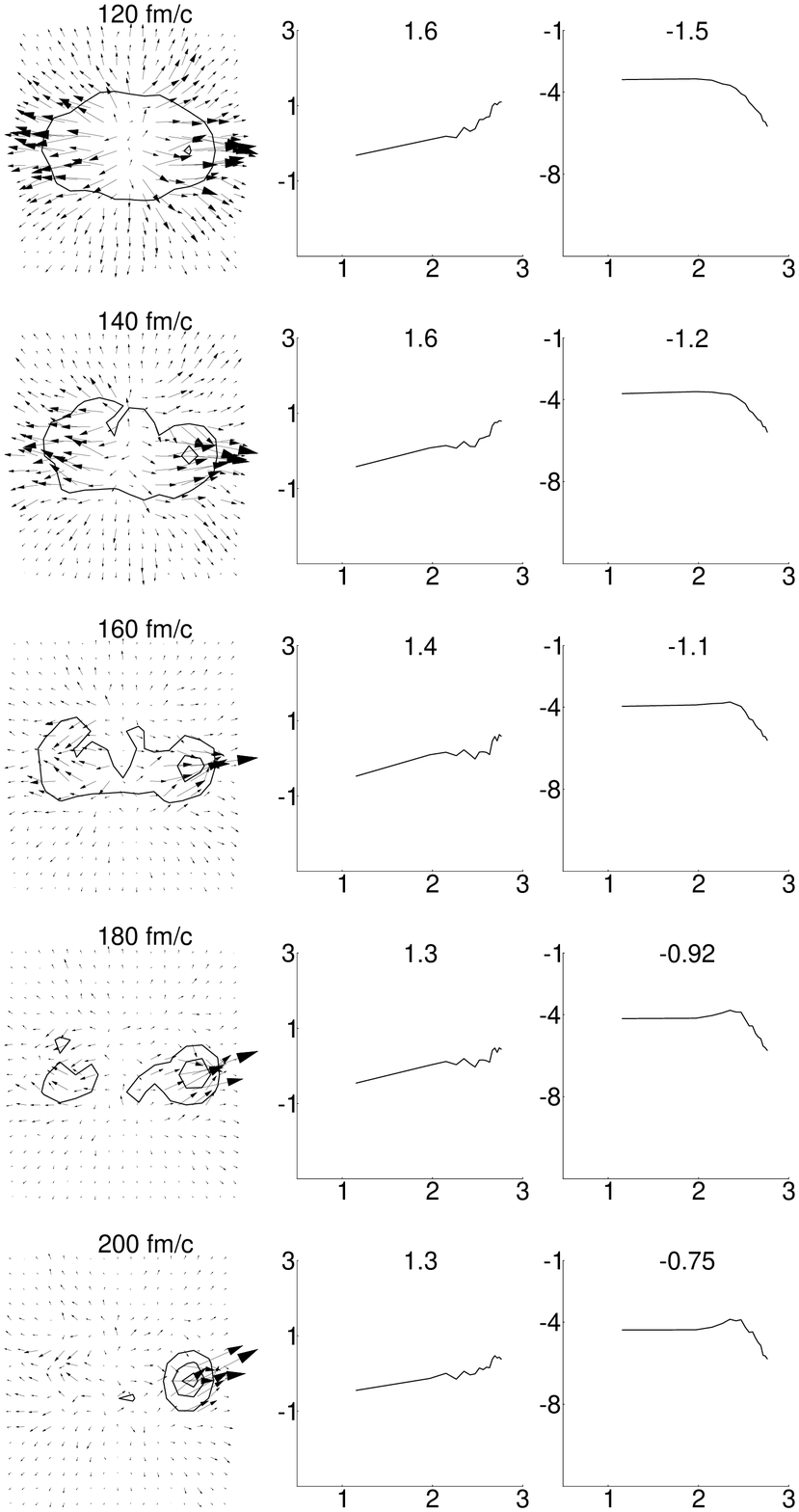,height=17cm,angle=0} 
\caption{The same as in Fig. 2 but for $E_{lab}/A = 60$ 
MeV.}
\label{fig2a} 
\end{figure}

The evolution picture shown in Figs. \ref{fig1} and 
\ref{fig1a} 
is essentially changed at higher bombarding energies. 
At $E_{lab}/A = 60$ MeV (see Figs. \ref{fig2} and 
\ref{fig2a}), 
the time interval where the Coulomb force 
counter-balance 
the nuclear forces is becoming very short ({\it e.g.}, 
see the plots for $t = 40$ fm/c) and the system
enters very fast in the phase of a smooth radial expansion. 
In this case, the velocity - radius scaling can be well approximated 
asymptotically by a single exponent 
$\alpha_{bulk}=\alpha_{surf} \simeq 1$. 
The particle density is  expanding almost uniformly and no characteristic 
power law dependence  is seen in the surface region. The 
proton excess in the surface region is less pronounced than at $E_{lab}/A = 33$
MeV/A (see Fig. 1) and $Z/N \simeq 1$ at later times. One expects
that  with increasing bombarding energy in symmetric HI reactions 
the effective source parameters for protons and neutrons become close to
each other.

\subsection{The expansion velocity}

\begin{figure} 
\psfig{file=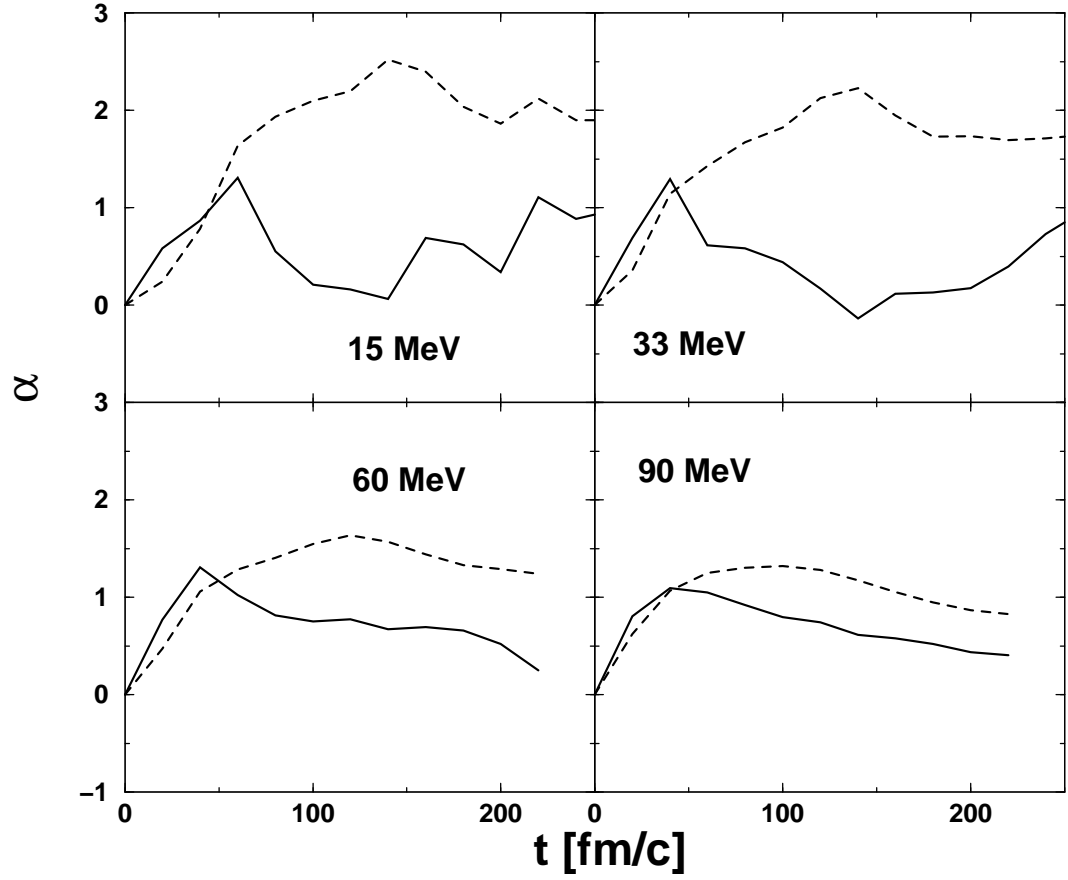,width=12cm,angle=-90}
\caption{The exponent $\alpha$ of the power law velocity 
profile (2) with 
  respect to the radius $v\propto r^\alpha$ for 
different lab
  energies. 
The dotted lines show the surface matter 
  behavior and the solid lines depict the bulk matter 
behavior.} 
\label{fig3} 
\end{figure}

Let us now discuss the dependence of the expansion 
velocity on 
the radius in more details. As was noted above,  
two different slopes can be distinguished at lower 
collision energies. 
We call the 'inner' and the 'outer' parts of the density 
profile 
the 'bulk' and the 'surface' regions, respectively. They 
are separated here 
at around $R = 10$ fm.  Therefore, we plot  in Fig. 
\ref{fig3} 
the time dependence of the exponent $\alpha$ in these 
two regions
for different bombarding energies.

Let us start with the reaction at $E_{lab}/A = 33 $ MeV.
The very first stage of the collision, where the nuclei 
are slightly overlapping, is characterized by similar 
values of 
the exponents $\alpha$ in the bulk and the surface 
regions. 
This feature continues until the surface particles begin
to be evaporated. At around this time, the
surface develops a much steeper velocity gradient with
$\alpha_{surf}$ as large as  $\sim 2$.  At the same 
time, the bulk matter velocity profile is quite smooth 
and even 
$\alpha_{bulk}$ changes the sign \cite{comment1}. 
After this overshooting of the surface exponent during a 
time interval
$100 - 200$ fm/c, the bulk matter develops 
the radial expansion with the coefficient 
$\alpha_{bulk}$ which is  
approaching $\alpha_{surf}$. Note, that even at this 
stage the mass current 
is characterized by the nonlinear scaling. Hence, using 
the Hubble 
ansatz for the radial flow in the analysis of
experimental data in the Fermi energy domain is not 
justified.

At lower energies ($E_{lab}/A=15$ MeV in Fig. 5), 
one  observes that the difference between radial flow 
patterns in the 
surface and bulk regions becomes even stronger than at 
$E_{lab}/A=33$ MeV. 
Consequently, higher values of the exponent 
$\alpha_{surf}$ 
are reached asymptotically. Moreover, we see that a 
giant resonance with
a period of $T=60$ fm/c, or an energy of $2 \pi/T=20.6$ 
MeV, is
excited. This  can be considered as a complete fusion 
event.

If one proceeds to energies higher than the Fermi 
energy, one sees
 that the general trend of evolution is conserved but
 the deviation between surface and bulk matter expansion 
patterns decreases, 
{\it i.e.}, the coefficients $\alpha_{bulk}$ and  
$\alpha_{surf}$ become close one to another. One can see 
also that 
the maximum value of the exponents $\alpha$ is reached 
faster 
than at lower energies, and 
the exponents $\alpha$ are becoming close to 1 at late 
times . 
The nuclear system at this high excitation energy is 
expanding continuously.
Due to the higher initial velocity, there is 
practically no inversion of the velocity profile  and no 
time-periodic 
structures like giant resonances are
excited. The Hubblean expansion pattern is reached 
faster and no anomalous
behavior ($\alpha_{surf} > 1$) is observed for energies 
higher than $90$ MeV/A.

\subsection{Dynamical trajectories}

To elucidate the connection of the anomalous velocity 
profile  with 
the multifragmentation, let us characterize an 
instantaneous  state of the 
system in terms of the average nucleon density $n$  and 
the temperature $T$.
In order to define a global time dependent temperature 
$T(t)$ we adopt the
Fermi liquid relation \cite{MT00} :
\be
{E}(t)=\frac 3 5 {E}_F(t)+ { E}_{\rm coll}(t) +
{\pi^2\over 4 { E}_F(t)}  \ { T(t)}^2  ~ \ ,
\label{t}
\ee
where the global kinetic energy $E(t)$, the collective 
energy $E_{\rm
  coll}$ and the Fermi energy $E_F(n)=(3 \pi^2 
n/2)^{2/3}/2m$ are given
by a spatial integration of the local quantities (3) :
\be
{E}(t)&=&{\displaystyle{\int d{r} \,{E}({
      r},t)}\over \displaystyle{\int d{r} \, 
{n}({r},t)}}
\nonumber\\
{E}_F(t)&=&{\displaystyle{\int d{r} \, {E}_F \left({n}({
    r},t)\right) \ {n}({r},t)} \over \displaystyle{\int 
d{r} \, {n}({r},t)}} \\
{E}_{\rm coll}(t)&=&{\displaystyle{\int d{r} 
\,{{J}({r},t)^2
    \over  m \,{n}({r},t)}} \over \displaystyle{\int 
d{r} \, {n}({r},t)}} 
\nonumber ~ \ .
\ee

It is more problematical to define a density. We will 
present here the two
possibilities. The first one is to consider the density 
of matter inside
the evolving mean square radius. The other possibility 
is to consider
the density of matter contained in the static separation 
of bulk
matter by the radius $R$ found earlier in the velocity 
profile.

The dynamical trajectories in the $(T-n)$ - plane for 
different collision energies are shown in Fig. 6 for the 
bulk matter ($R < 10$ fm) . Let us
first look at the static density definition 
inside the bulk region. For $E_{lab}/A=$ 15 and
33 MeV, the system evolves inside the spinodal region 
and, 
therefore, is mechanically unstable. In the time 
interval between 150 fm/c
and 200 fm/c, the system is in a configuration with 
$n\approx n_0/3$ and $T\approx 7-8$ MeV,
which are the typical values for the nuclear 
multifragmentation. 
At higher energies, the freeze-out density of an
evolving system is shifted towards lower densities and 
finally it ends
in a gaseous phase. If one chooses to look at 
the interior part of the system within the radial size 
of 
the mean squared radius (dotted lines in Fig. 6), the 
main difference is seen
in the initial stage of the evolution where larger 
nucleon densities are 
reached but, at the same time, they are passed through 
much faster. 
The freeze-out configurations are practically the same 
as when the 
above static definition of bulk matter is considered.

\subsection{The long tail of the density distribution}
Let us now discuss the particle density profile in more
details. As noted in Fig. 2, the decrease of the 
particle 
density in the surface region is algebraic (power law) 
rather than exponential. In Fig. 7, we plot the
time dependence of the exponent $\beta$ which is 
extracted from the power law fit : $n(r) \sim 
r^{-\beta}$, for $R > 10$ fm. 

There are here two distinct behaviors. For $E_{lab}/A =
15$ and 33 MeV, after an initial build-up of the surface 
region as
characterized by the decrease of $\beta$ in time, the 
value of exponent $\beta$
stabilizes asymptotically. The limiting value of 
$\beta$, called $\beta_{lim}$,
decreases with
bombarding energy from $\beta_{lim} \sim 3.5$ at 
$E_{lab}/A = 15$ MeV 
to $\beta_{lim} \sim 3.1$ at $E_{lab}/A = 33$ MeV. 
Actually, the most 
interesting long-range 
region of the nucleon density ends at about 
$E_{lab}/A \sim 50$ MeV and the smallest value reached is about
$\beta_{lim}=2$. In the energy interval 
$15$ MeV $\lsim E_{lab}/A \lsim 50$
MeV, the anomalous power law tail of the nucleon density 
accompanies the anomalous profile of the expansion 
velocity in the surface
region, and both effects are leading to the 
non-Gaussian shape of the emitting source and the
anomalous short-range correlations. 

For energies $E_{lab}/A \geq 60$ MeV in the Hubblean expansion regime, 
the exponent $\beta$ decreases monotonously in time 
and no asymptotically stable surface region with
the power law dependence is seen. At later times , one sees however the
appearance of a new expansion regime corresponding to the 
negative values of the exponent $\beta$. This indicates
the formation of a shell-like structure in the system. The formation
time of the shell-like structure decreases rapidly with increasing bombarding
energy ($t_{sh} \simeq 280$ fm/c at 
$E_{lab}/A = 60$ MeV and  $t_{sh} \simeq 170$ fm/c at $E_{lab}/A = 90$ MeV).
The $\beta$ values found are not strongly negative, 
indicating that the expanding shell
is a very diffused object due to the  
long mean-free path of nucleons in the kinetic approach. Similar
unusual solution of spherically expanding 
scaling hydrodynamic has been used in the
analysis of the Bose-Einstein correlations \cite{Cs98}.

The specific shape fluctuations in the long-range 
region can be a source of the
power-law Bose-Einstein correlations. This problem has 
been discussed by
Bia{\l}as \cite{bia,rep} in the context of strong 
interaction physics at
relativistic energies. The long tail of the particle 
density for the systems
produced in the symmetric heavy-ion collisions in the 
Fermi energy domain, 
may result in the unusual quantum statistical 
correlations. 
The $N$-particle interferometry reduced densities can be 
written as :
\begin{eqnarray}
\label{b1}
&&D_N(k_1,\dots ,k_N)= \nonumber \\&=&\int dx_1\dots 
\int dx_N\left(
\sum_{per}\exp[i(x_1p_{a_1}+\dots +x_Np_{a_N})]\right) 
\times \nonumber
\\&\times& n_N(x_1,\dots
,x_N) \left[{N!\left( \int dx 
n_1(x)\right)^N}\right]^{-1} ~ \ ,
\end{eqnarray}
where $n_N$ is the $N$-point density of emitting sources 
and the sum runs over
all permutations of the indices $a_i$. The above formula 
supposes the
incoherent emission from the source and neglects the 
final state interaction. In the case of
uncorrelated emission in space-time :
$$n_N(x_1,\dots ,x_N)=n_1(x_1) \cdots n_1(x_N)~ \ ,$$ 
the $N$-particle cumulant can be written using the 
Fourier
transform of the source density $n_1(x)$ \cite{bia} :
\begin{eqnarray}
\label{b2}
n_N(k_1,\dots ,k_N)=\sum n_1(k_1-k_{a_1}) \cdots 
n_1(k_N-k_{a_N}) ~ ,
\end{eqnarray}
where the sum runs over all permutations of the indices 
$a_i$ with $a_i \neq
i$. If the source density has a power law tail :
\begin{eqnarray}
\label{b3}
n_1(x)\simeq x^{-\beta} \equiv x^{\gamma - D} \ ,
\end{eqnarray}
its Fourier transform shows also a power law in some 
range of small momenta
\cite{bia} :
\begin{eqnarray}
\label{b4}
n_1(k)\simeq |q|^{-\gamma} \ .
\end{eqnarray}
Thus there is the relation between the power law tail in 
the source density
distribution and
the power law in the two-particle Bose-Einstein 
correlations which are given
in terms of the Fourier transform of the source density 
:
\begin{eqnarray}
\label{b5}
C_2(q) \simeq |q|^{-2\gamma} ~ \ .
\end{eqnarray}
Similarly, the higher order cumulants $C_i(q)$ 
($i=3,\dots$)
are expected also to have a power law
dependence on the rescaling of momenta with an index 
$i\gamma$. 

A particularly interesting case of the Bose-Einstein 
correlations, increasing
as the power of $|q|$, corresponds to $\gamma > 0$ 
($\beta < D$) in
Eq. (\ref{b3}). In the case of central heavy-ion 
collisions, our analysis suggests
that this effect could be seen in symmetric systems for 
energies : 
$35$ MeV $\lsim E_{lab}/A \lsim 50$ MeV.

\section{The summary and outlook}

The dynamical behavior of heavy-ion reactions in the 
Fermi energy domain during the
first $200$ fm/c is clearly associated with a nonlinear 
$(\alpha > 1)$ 
radial velocity profile (2). The existence of such a 
non-Hubblean radial
expansion at the freeze-out configuration 
was postulated by Le F\'{e}vre et al. \cite{LPT99} in 
the analysis of 
experimental kinetic energies of fragments in the $Xe + 
Sn$ reaction at 
$E_{lab}/A = 50$ MeV  in the 
framework of the statistical
microcanonical model. The present studies using the 
nonlocal
kinetic theory show that indeed such an unusual radial 
flow velocity
profile is a plausible freeze-out configuration 
for symmetric heavy-ion collisions.
One can roughly determine the instant of time when the 
compression 
turns into the expansion by looking at the crossing 
point of the bulk  
$\alpha_{bulk}$ and surface  $\alpha_{surf}$
scaling exponents. While at the compression stage 
these two exponents are almost equal, the expansion 
stage exhibits 
larger exponent for the surface matter than 
for the bulk matter and $\alpha_{surf}$ 
takes values significantly larger than 1. This dynamical
behavior of the surface matter disappears for energies 
significantly higher 
than the Fermi energy. 

For central heavy-ion collisions in the Fermi energy 
domain, 
we see that the expansion stage is characterized by very 
small values 
of the exponent $\alpha$ in the bulk 
($\alpha_{bulk} \in [0,1]$), indicating its slow 
evolution
and possible  mechanical instabilities. At these 
energies, 
the system spends a long time in the
spinodal region, what may result in its 
multifragmentation decay.
With increasing collision energy, the system passes 
quickly through 
this unstable region or, at even higher energies, its 
thermodynamic 
trajectories go above the spinodal region. In this case, 
the 
multifragmentation due to the spinodal instabilities is 
hardly possible
\cite{comment2}. For details see \cite{MT00}

The nonlinear radial velocity profile in the surface 
region is
accompanied by the long tail of the particle density. 
Both these effects may
have important consequences on the quantum statistical 
correlations and their
evolution with the bombarding energy. The effect of the 
radial expansion on the
two-particle correlations have been studied assuming the
linear scaling solution ($\alpha = 1$) of the scaling
hydrodynamics \cite{CLZ94,HCA97}. It was found that
the expansion makes the effective radius of the 
two-particle correlation
functions smaller than the geometrical size of the 
source. We expect this
effect to be even stronger in the presence of the 
nonlinear radial 
expansion flow, leading to even stronger discrepancy 
between the effective
radius and the geometrical radius. Moreover, the 
commonly used Gaussian
approximation for the source shape is certainly 
hazardous in this 
non-Hubblean expansion regime. The existence of the 
power law tail in the
particle density can in turn lead to the power law 
two-particle Bose-Einstein
correlations at small relative momenta of the particles in the range of
collisions energies $35$ MeV $\lsim E_{lab}/A \lsim 50$ MeV. It should be
stressed that the ratio $Z/N$ in the surface region is strongly different from
1, in particular at early collision times and at 
low bombarding energies ($E_{lab}/A \lsim 60$ MeV). This specific
effect in 
the Fermi energy domain, 
which leads to an
effective increase of the proton rms radius  and to the
change in the temperature gradients in proton/neutron source, may
have measurable consequences in the quantum statistical
correlations for protons and neutrons. This effect has been studied by
Helgesson et al. \cite{HCA97} assuming a linear scaling solution 
($\alpha = 1$) of hydrodynamics for asymmetric HI reactions at $30$
MeV/A. Our results suggest that the simultaneous description of $n$ and $p$
spectra, as well as $nn$ and $pp$ correlation functions may require different
source parametrizations for neutrons and protons, though the detailed dynamics
for HI reactions at $E_{lab}/A \simeq 30$ MeV is very different from
those assumed by Helgesson et al. \cite{HCA97}. It is interesting to notice
, that the nonlocal kinetic theory predicts the appearance of the 
solution somewhat similar to the linear scaling solution ($\alpha = 1$) 
of hydrodynamics with $\beta < 0$ \cite{Cs98} 
for higher energies bombarding energies ($E_{lab}/A \gsim 60$ MeV.

The different behavior of the surface and bulk matter 
before
equilibration can also be of importance for the 
description of super nova
where a surface - like ring of matter (crust) 
is expanding with enormous velocities and
is clearly separated from the remaining bulk matter 
collapsing back into
neutron stars. The clearest experimental 
observation of the expanding
shell-like structure noted above comes from stellar 
astronomy. The envelope
material ejected by the stars forms an expanding shell 
of gas that is known as
a planetary nebula. The space-time evolution of these 
objects is in many
aspects similar to the considered evolutions :  $\alpha \lsim 2$, $\beta > 0$ 
and  $\alpha \sim 1$, $\beta < 0$. The latter solution can be
successfully simulated by a
scaling solution of the non-relativistic hydrodynamics \cite{Cs98}.  

We suggest that the anomalies found in the kinetic 
expansion 
reflect the nature of effective interactions among the 
elementary constituents of the system. Their  
manifestation is twofold.
Firstly, the interplay between the repulsive Coulomb 
interaction 
and the attractive mean field results in the formation 
of a rather
sharp  surface of the system. Secondly, the evolution of 
the bulk 
matter is not a simple uniform expansion of a 
homogeneous ideal fluid.
In the unstable spinodal region, the local interaction 
of 
quasiparticles in the bulk phase affects noticeably the 
subsequent evolution of the system. In this respect, one 
should
stress again an attractive possibility of implementing 
both the
nonlinear velocity profile and the algebraic long-range 
density tail 
into the analysis of the Bose-Einstein correlations. 
These effects are
not only important for the nuclear multifragmentation 
process but also for the
hadron interferometry at ultrarelativistic collisions 
where the deconfinement 
phase transition can have a strong influence on the 
expansion stage. This is
of a particular interest for the statistical mixed phase 
equation
of state which gives the crossover type of the 
deconfinement
phase transition and allows for a small admixture of 
unbound quarks
at the freeze-out point \cite{NST98}.     
  
An interesting example which illustrates an
important difference between the linear and quadratic 
scalings of the
expansion velocity with the radius can be found in 
cosmology \cite{H73}.
If there is an attractive force decreasing as an inverse 
power of the
radius,  the equation of motion
for a radially symmetric matter follows from the total 
energy $h$ which reads :
\be
\frac m 2  {\dot R(t)}^2- {G \over R(t)^\delta}= h ~ \ .
\ee
Assuming a homogeneous matter density $n$, the mass is~:
$m=4\pi n R^3/3$. In the case of the escaping matter 
$(h=0)$, one gets :
\be
{\dot R(t)}=\sqrt{{6 G \over 4 \pi n}} \ 
R(t)^{\displaystyle {3-\delta\over 2}} ~ \ ,
\ee
what corresponds to taking :  
\be
\alpha=\frac{3-\delta}{2}
\ee 
in Eq. (\ref{eq2}). We see that $\delta=1$ for the 
Coulomb or gravitational 
forces and, hence, the Hubble expansion ($\alpha = 1$) 
follows in these cases. However if $\alpha=2$, 
as found in the surface region of nuclei formed in 
central HI
collisions at around the Fermi energy (see Fig. 5), the 
above relations 
lead to a string-like force with $\delta=-1$. 
Since this force is used to describe the confinement in
the effective theories motivated by the Quantum 
Chromodynamics, 
its occurrence as a consequence of the {\em dynamical 
behavior} is worth of attention. As was shown above 
in the solution of nonlocal kinetic
equations, the interplay between Coulomb and mean field 
leads to such a
string-like behavior for surface particles.
This is here a clear non-equilibrium effect. The above 
discussed 
oscillation in the time-dependence of the $\alpha$ 
exponent may be
considered as a possible manifestation of this effective 
 dynamical
string-like force in the surface region.

Perhaps the best way to demonstrate the existence of 
both the 
non-Hubblean radial expansion and the algebraic 
long-range tail of the particle
density, would correspond to finding a non-Gaussian 
deformation of
the source in the Bose-Einstein interference experiments 
for central, symmetric 
heavy-ion reactions in the narrow range of
collisions energies ($35$ MeV $\lsim E_{lab}/A \lsim 50$ 
MeV). In the same
narrow range unusual scale - dependence of the 
many-particle
correlations at small momenta should be induced via the 
same Bose-Einstein quantum
interference effect if the source has an algebraic 
(power law) 
density tail (Sect. II.E). On the other hand, as 
discussed in \cite{LPT99}, 
the kinematical observables related mainly to the 
intermediate mass fragments,
provide an independent and sensitive measure of 
the velocity profile in the deformed, expanding source. 
To put
together all these different pieces of evidence into the 
circumstantial proof
for the non-Hubblean collective expansion of nuclear 
aggregates, 
remains a difficult and exciting challenge
for the future experimental and theoretical studies 
of heavy-ion collision in the Fermi energy domain.

\section{Acknowledgements }
This work was supported by the IN2P3-JINR agreement No 
0049. K. M. likes to thank the LPC for hospitality and 
friendly atmosphere.

\onecolumn 
\vspace*{1cm} 

\newpage

\newpage 

\newpage 

\newpage 
\vspace*{1cm} 

\newpage
\begin{figure} 
\psfig{file=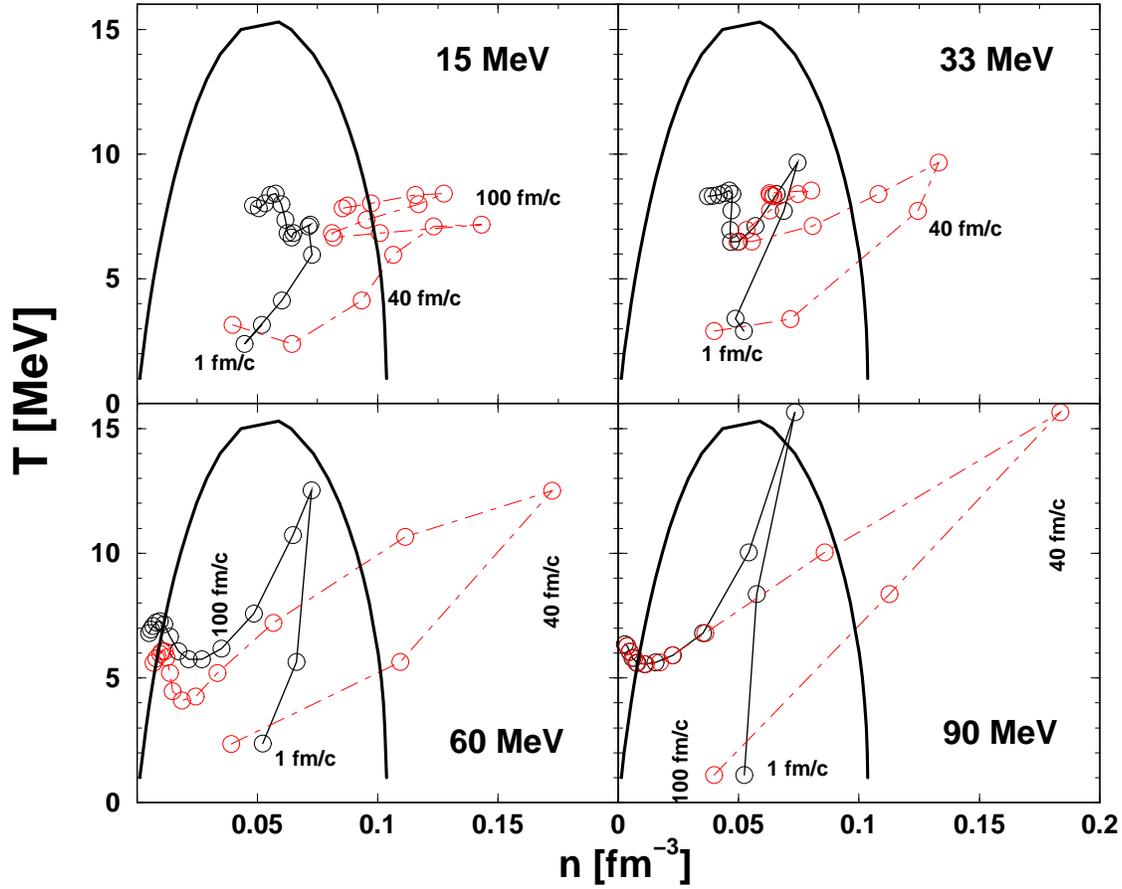,width=12cm,angle=-90}
\caption{The time evolution of the temperature versus 
the density for 
the sphere of a constant radius ($R < 10$ fm) (the solid 
line) corresponding to 
the distinction between the bulk and surface regions in 
Fig. 1, and for
  the sphere given by the mean square radius of 
  the evolving  system  (the dotted line). 
The thick solid lines give the limit of the spinodal 
  region for an infinite nuclear matter.} 
\label{fig4} 
\end{figure}

\newpage 
\vspace*{1cm} 

\begin{figure} 
\psfig{file=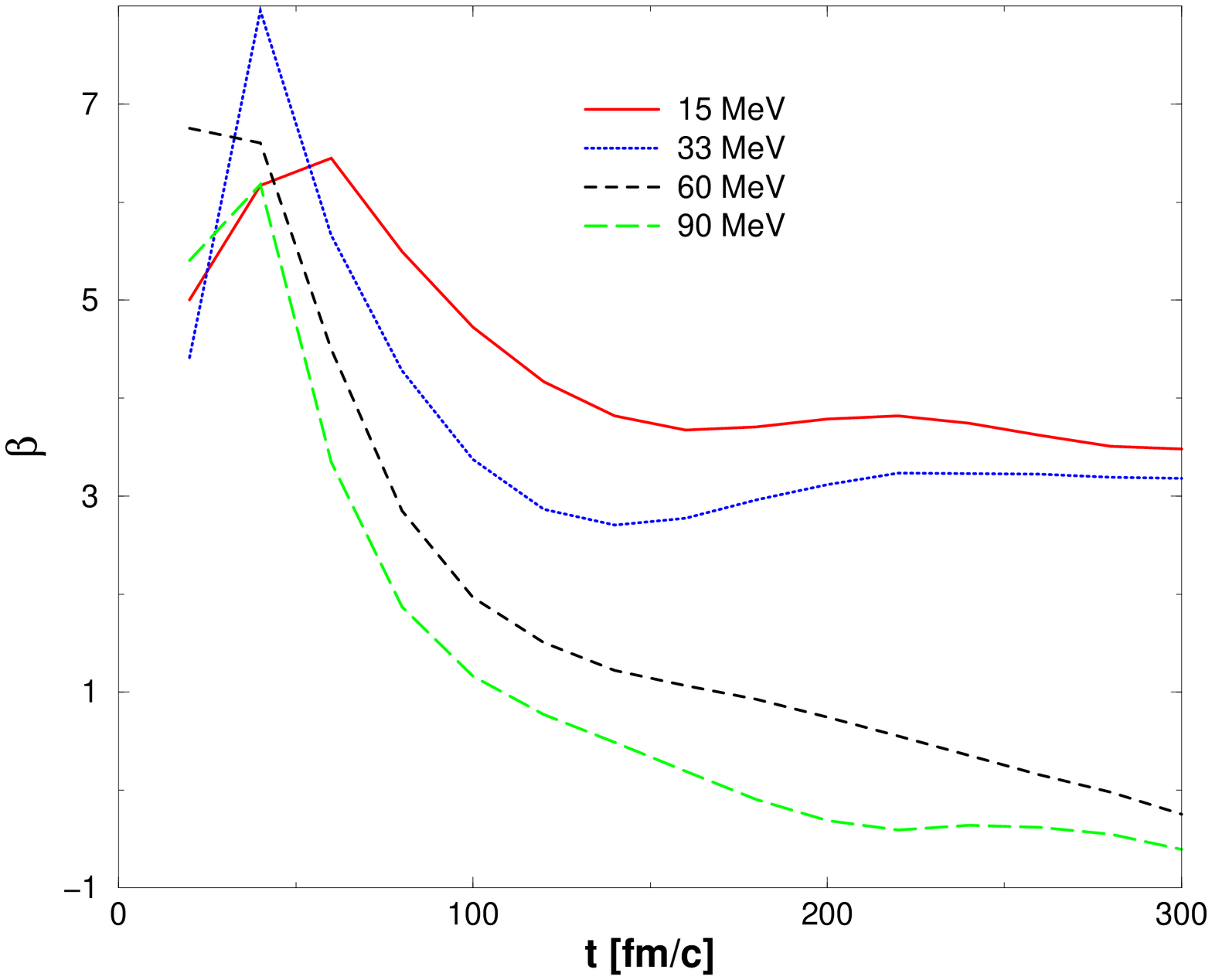,width=12cm,angle=-90}
\caption{The time dependence of the exponent $\beta$ of 
the power law fit of the particle density $n\propto 
r^{-\beta}$ in the surface region
($R > 10$ fm) for different collision energies.}
\label{fig7}
\end{figure}

\twocolumn 

\begin{thebibliography}{50}

\bibitem{Dan99} P.~Danielewicz, Nucl. Phys. A {\bf 661}, c82 (1999).
 
\bibitem{MS88} A.N.~Makhlin and Y.M.~Sinyukov, Z. Phys. 
C {\bf 39}, 69 (1988).

\bibitem{CLZ94} T.~Cs\"{o}rgo, B.~L\"{o}rstad, and 
J.~Zim\'{a}nyi, Phys. Lett. B {\bf 338}, 134 (1994);\\
T.~Cs\"{o}rgo and B.~L\"{o}rstad, Phys. Rev. C {\bf 54}, 
1390 (1996);\\
B.~Tom\'{a}sik and U.~Heinz, Eur. Phys. J. C {\bf 4}, 
327 (1998).

\bibitem{PMB98} A.~Polleri, I.N~Mishustin, J.P.~Bondorf,
   Phys.Lett. B {\bf 419}, 19 (1998).

\bibitem{BGZ78} J.P.~Bondorf, S.I.A.~Garpman, and 
J.~Zim\'{a}ny, Nucl. Phys. A {\bf 296}, 320 (1978).

\bibitem{CCL98} P.~Csizmadia, T.~Cs\"{o}rgo, and 
B.~Luk\'{a}s, Phys. Lett. B {\bf 443}, 21 (1998).

\bibitem{H73}
M. Harwit, {\em Astrophysical Concepts} (John Wiley{\&}Sons, 
New York, 1973).

\bibitem{MTW72} Ch.W.~Misner, K.S.~Thorne, and 
J.A.~Wheeler, 
{\em Gravitation}, W.H.~Freeman and Co, San Francisco, 
1972.


\bibitem{D99} A.~Dumitru,  Phys. Lett. B {\bf 463}, 138 
(1999).

\bibitem{LPT99} A.~Le~F\'{e}vre, M.~P{\l o}szajczak and 
V.D.~Toneev, Phys. Rev. C {\bf 60}, R051602 (1999). 

\bibitem{SLM96}
V. {\v S}pi{\v c}ka, P. Lipavsk{\'y}, and K. Morawetz, Phys. Lett. 
A {\bf 240}, 160 (1998).

\bibitem{LSM99}
P. Lipavsk{\'y}, V. {\v S}pi{\v c}ka, and K. Morawetz, 
Phys. Rev. E {\bf 59},
R1291 (1999).

\bibitem{NTL91}
P.~J. Nacher, G. Tastevin, and F. Laloe, Ann. Phys. 
(Leipzig) {\bf 48},  149
  (1991).

\bibitem{H90}
M. de~Haan, Physica  A {\bf 164},  373  (1990).

\bibitem{MLSK98}
K. Morawetz, P. Lipavsk{\'y}, V. {\v S}pi{\v c}ka, and 
N. Kwong, Phys. Rev. C {\bf 59},  3052  (1999).

\bibitem{MLSCN98}
K. Morawetz {\it et~al.}, Phys. Rev. Lett. {\bf 82}, 3767  (1999).

\bibitem{HCA97}
J. Helgesson et al., Phys. Rev. C {\bf 56}, 2626 (1997).

\bibitem{BRRS93}
B. Borderie, B. Remaud, M. Rivet, and F. Sebille, Phys. Lett. B {\bf 302},  15
  (1993).

\bibitem{Dan95}
P. Danielewicz, Phys. Rev. C {\bf 51},  716  (1995).

\bibitem{comment1}
We have checked that such a bouncing back effect is 
more pronounced in the standard BUU calculations
than in the nonlocal kinetic scenario.

\bibitem{MT00}
K. Morawetz, nucl-th/0004024, Phys. Rev. C 62, 0446xx (2000).

\bibitem{Cs98}
 T.~Cs\"{o}rgo,  nucl-th/980911, hep-ph/0001233. 
\bibitem{bia}
A. Bia{\l}as, Acta Phys. Pol. B {\bf 23}, 561 (1992);\\
A. Bia{\l}as and B. Ziaja, Acta Phys. Pol. B {\bf 24}, 
1509 (1993).

\bibitem{rep}
P. Bo\.zek, M. P{\l}oszajczak and R. Botet, Phys. Rep. {\bf 252}, 101 (1995).

\bibitem{comment2}
 This does not exclude certainly that the system can be 
in 
the spinodal instable region for larger impact 
parameters.


\bibitem{NST98} E.G.~Nikonov, A.A.~Shanenko, and 
V.D.~Toneev,
Heavy Ion Physics {\bf 8}, 89 (1998);\\
 V.D.~Toneev, E.G.~Nikonov, and A.A.~Shanenko,
in {\it Nuclear Matter in Different Phases and 
Transitions}, Eds. J.-P.~Blaizot,
X.~Campi, and M.~P{\l}oszajczak, Kluwer Academic 
Publishers (1999), p. 309.




\end{thebibliography}
\end{document}